# Perfect energy-feeding into strongly coupled systems and interferometric control of polariton absorption

Simone Zanotto[1], Francesco P. Mezzapesa[2], Federica Bianco[1], Giorgio Biasiol[3], Lorenzo Baldacci[1], Miriam Serena Vitiello[1], Lucia Sorba[1], Raffaele Colombelli[4], and Alessandro Tredicucci[1]

[1]*NEST, CNR-Istituto Nanoscienze and Scuola Normale Superiore, Piazza San Silvestro 12, I-56127 Pisa (Italy)*

[2]*CNR-Istituto di Fotonica e Nanotecnologie UOS Bari and Dipartimento Interateneo di Fisica, Università degli Studi di Bari "Aldo Moro", Via Amendola 173, I-70126 Bari (Italy)*

[3]*Laboratorio TASC, CNR-IOM, Area Science Park, S.S. 14 km 163.5 Basovizza, I-34149 Trieste (Italy)*

[4]*Institut d'Electronique Fondamentale, Univ. Paris Sud, UMR8622 CNRS, F-91405 Orsay (France)*

**The ability to feed energy into a system, or – equivalently – to drive that system with an external input is a fundamental aspect of light-matter interaction. The key concept in many photonic applications is the *"critical coupling"* condition [1,2]: at criticality, all the energy fed to the system *via* an input channel is dissipated within the system itself. Although this idea was crucial to enhance the efficiency of many devices, it was never considered in the context of systems operating in a non-perturbative regime. In this so-called *strong coupling regime*, the matter and light degrees of freedom are in fact mixed into *dressed states,* leading to new eigenstates called *polaritons* [3-10].**
Here we demonstrate that the *strong coupling regime* and the *critical coupling condition* can indeed coexist; in this situation, which we term *strong critical coupling,* all the incoming energy is converted into polaritons. A semiclassical theory – equivalently applicable to acoustics or mechanics – reveals that the *strong critical coupling* corresponds to a special curve in the phase diagram of the coupled light-matter oscillators. In the more general case of a system radiating via two scattering ports, the phenomenology displayed is that of *coherent perfect absorption* (CPA) [11,12], which is then naturally understood and described in the framework of critical coupling. Most importantly, we experimentally verify *polaritonic CPA* in a semiconductor-based intersubband-polariton photonic-crystal membrane resonator.

This result opens new avenues in the exploration of polariton physics, making it possible to control the pumping efficiency of a system almost independently of its Rabi energy, i.e., of the energy exchange rate between the electromagnetic field and the material transition.

In opto-electronic devices energy dissipation can be highly undesirable or very much needed, depending on the foreseen application. Photon detectors and solar cells are the prototypical semiconductor components belonging to the second category: the incoming electromagnetic energy must be mostly transferred and dissipated within the active region for optimized operation [13]. The same condition holds for most microwave and radio systems: an antenna must efficiently deliver/extract the signal to/from the receiving/emitting circuit. The crucial concept is here the *critical coupling* between input channel and load. It is nothing else than the so-called impedance matching condition, which, once expressed in terms of losses, indeed states that radiative and material losses must be equal [1]. On one hand the *critical coupling* provides a common framework to the efforts aiming at developing perfect absorbers and optimal thermal emitters [14]. On the other hand, it is noticeably similar to the lasing condition, upon exchanging *losses* with *gain*. Indeed, the link between perfect absorbers and lasers has been recently pointed out in terms of *time reversal* within the coherent perfect absorption framework (CPA) [11].

In all the aforementioned examples, the material loss mechanisms (resistive, impurity scattering, interband electronic transitions…) feature a much larger bandwidth than the energy exchange rate with the electromagnetic field. Therefore, independently of the presence of a resonator and its radiative losses, light-matter interaction can be described under a perturbative approach, simply in terms of absorption and spontaneous emission processes. On the contrary, when a material excitation (an atomic transition, for instance) is coupled to an optical resonator having a similar bandwidth and a sufficiently small modal volume, the *strong coupling regime* occurs [15]. In this situation new quantum eigenstates of the system are formed as a combination of material and photonic excitations (polaritons) and energy is continuously exchanged between the two "fields" at a rate corresponding to the coupling strength (the so-called vacuum Rabi energy). One question now naturally arises: is it possible to feed perfectly and in full the electromagnetic energy from the outside world into such mixed light-matter polariton states? Or, in other words, is it possible to "critically couple" polaritons without destroying them?

A rigorous description of dissipation in a quantum picture would require a master equation approach [16 - 21], or a Green's function formalism [22]. However, under the hypothesis of a small average excitation density, a simpler semiclassical approach can be employed. The matter degrees of freedom *and* the cavity photons are represented as coupled harmonic oscillators, and a coupled-mode theory - which is developed in this paper – yields analytical formulas equivalent to those reported in the above mentioned literature. This analytical formalism reveals that the phenomenon of CPA - in either weak- or strong-coupling regime - is more deeply understood in terms of matching of damping rates: no additional parameters need to be tailored. Furthermore, the relevance of strong critical coupling rather than, for example, perfect transmission [23], is motivated by the need of efficient optical pumping in the proposal of various intersubband-polariton devices [24,25]. There, maximizing the absorption into polariton states along with tailoring the Rabi splitting is of crucial importance, and the concept of strong critical coupling enables to independently tune these two quantities that would otherwise be simultaneously controlled by a single parameter (i.e., the charge density).

Coupled-mode theories (CMT) are commonly applied to deal with different problems involving stationary and propagating modes in optics [1,2]. In the following, we will employ the notation introduced for photonic crystal resonances (Dirac notation) [26]. The system under study is sketched in Fig. 1 (a): it consists of an optical resonator, modeled as two partially reflecting mirrors, driven from the exterior via a coupling constant $d$. The system is assumed to be spatially symmetric (the effect of asymmetry will be discussed in a forthcoming publication). The optical cavity resonance occurs at a pulsation $\omega_0$ and it coherently exchanges energy with a matter resonator oscillating at the same frequency via the coupling constant $\Omega$. Energy is also re-radiated towards the exterior. The amplitudes of the cavity and matter fields ($a$ and $b$ respectively), the amplitudes of the incoming waves in the two ports $|s^+\rangle = (s_1^+, s_2^+)$, and the corresponding outgoing wave amplitudes $|s^-\rangle$, are then related by the following equations:

$$\frac{db}{dt} = (i\omega_0 - \gamma_m)b + i\Omega a$$
$$\frac{da}{dt} = (i\omega_0 - \gamma_c)a + i\Omega b + \left(\langle d|^*\right)|s^+\rangle \quad (1)$$
$$|s^-\rangle = C|s^+\rangle + a|d\rangle$$

The evolution of the matter and cavity oscillators is damped by the presence of two decay channels, quantified respectively with the decay rates $\gamma_m$ and $\gamma_c$. The latter is the cavity *total* damping rate, which is useful to separate into *radiative* and *non-radiative* contributions: $\gamma_c = \gamma_r + \gamma_{nr}$; $\gamma_r$ represents the radiative losses and $\gamma_{nr}$ keeps into account non-radiative, non-resonant cavity losses (for instance, free-carrier absorption and ohmic losses).

In steady-state the system response is given by the frequency-dependent scattering matrix which connects ingoing to outgoing waves: $|s^-\rangle = S(\omega)|s^+\rangle$. Straightforward integration of the above equations yields:

$$S(\omega) = C - \frac{i(\omega - \omega_m) + \gamma_m}{(\omega - \omega_-)(\omega - \omega_+)} D. \quad (2)$$

The explicit expressions for matrices $C$ and $D = |d\rangle(\langle d|)^*$ can be found in Ref. [26], while the polariton poles are

$$\omega_\pm = \omega_0 + \left[i(\gamma_c + \gamma_m) \pm \sqrt{4\Omega^2 - (\gamma_c - \gamma_m)^2}\right]/2.$$

Consistently with previous reports [27], this model predicts reflectance and transmittance lineshapes belonging to a Fano-like manifold; this complexity is lost, however, when dealing with the absorption properties, which are the object of the present work. It turns out then that a single spectral function is involved: $B(\omega) = (1 - |\det S(\omega)|^2)$. The explicit expression of the S-matrix determinant following from (2) is given in the Supplementary Information.

In a two-port system, the absorption is controlled by the simultaneous presence of two input beams; the relevant quantity is the joint absorbance $A_{joint}$, defined as the ratio between absorbed and input energies when the two coherent beams excite the system's ports with equal intensity. By sweeping the input beam dephasing $\varphi = \arg(s_2^+/s_1^+)$, the joint absorbance sweeps from a minimum $A_{joint, min}$ to a maximum $A_{joint, max}$, which in the CMT model are given by

$$A_{joint, min}(\omega) = 0; \quad A_{joint, max}(\omega) = B(\omega). \quad (3)$$

Indeed, if $\det S(\omega) = 0$ one has $A_{joint, max}(\omega) = 1$, i.e. a $\varphi$ can be found producing CPA. In addition, $A_{joint, min}(\omega) = 0$ means that it always exists another $\varphi$ which gives coherent perfect transparency (CPT). In essence, a device implementing the two-oscillator CMT with a S-matrix satisfying $\det S = 0$ will behave as an ideal absorption-based interferometer.

If, on the contrary, a single beam excites the system either from port 1 or 2, the usual single-beam absorbances are observed. From the CMT it follows that in this case $A_1(\omega) = A_2(\omega) = B(\omega)/2$; this is consistent with the general theory of coherent absorption that states that the average between single-beam absorbances coincides with the average between minimum and maximum joint absorbances (see Supplementary Information). This means, from one hand, that a sample whose single-beam absorption peaks at ½, is likely to show CPA; on the other hand, single-beam absorption never exceeds ½ in a symmetric resonator supporting *one* resonant electromagnetic mode, as predicted by previous CMT calculations [28].

The function $B(\omega)$ can exhibit either one or two peaks, depending on the region of the parameter space involved, as shown in Figure 1 (b); this transition can be regarded as a crossover between weak and strong coupling regimes. In both weak- and strong-coupling regions a curve in the parameter space exists corresponding to $\det S(\omega) = 0$, hence CPA (see Supplementary Information). For small enough matter damping rate ($\gamma_m < \Omega$), one observes CPA when the damping rate matching

$$\gamma_r = \gamma_{nr} + \gamma_m \quad \text{(strong critical coupling)} \quad (4)$$

is fulfilled; notice the contribution of non-radiative losses which can "help" a "good" matter resonator entering the strong critical coupling [29]. On the other hand, the same system supports a second kind of *critical coupling*, which occurs when the following is satisfied:

$$\gamma_m (\gamma_r - \gamma_{nr}) = \Omega^2 \quad \text{(weak critical coupling).} \quad (5)$$

In both weak and strong critical coupling all the incoming energy can be perfectly absorbed by the system, although it is obviously redistributed between the non-radiative cavity losses and the matter oscillator. A special case is found when the resonator-oscillator coupling $\Omega$ is set to zero: strong critical coupling no longer exists. Instead, the model shows that weak critical coupling reduces to the conventional CPA. This result is non-trivial, and it is also aesthetically gratifying: CPA is nothing else than critical coupling in a two-port system, and it obeys the simple, usual critical coupling condition $\gamma_r = \gamma_{nr}$. This situation is shown in Fig. 1 (c), where the absorption lineshapes given by $B(\omega)$ become purely Lorentzian.

To experimentally demonstrate strong critical coupling and full interferometric control of absorption, we performed an optical measurement on a polaritonic sample that mostly satisfies the radiative decay rate matching of Eq. (4). The CPA setup is sketched in Fig. 2 (a): the light source is a commercial external-cavity tunable quantum-cascade laser (Daylight Solutions) operating in the range of wavelengths 9.9 μm – 10.7 μm, the phase-delay stage is a loudspeaker membrane driven by a function generator, and the detectors are liquid-nitrogen cooled mercury-cadmium-telluride devices. The sample consists of a semiconductor membrane which has been structured as a photonic crystal slab resonator (Fig. 2 (b)); it embeds a multi-quantum well (MQW) heterostructure whose quantum design results in a single intersubband transition resonant with the photonic crystal mode, and hence in intersubband polariton states [30]. In detail, we employed a stack of 50 $Al_{.33}Ga_{.67}As/GaAs$ QWs, grown by molecular beam epitaxy, with barrier and well thicknesses of 30 nm and 8.3 nm, respectively, and a nominal sheet doping n = $5 \cdot 10^{11}$ cm$^{-2}$ in the well material. The resulting 2 μm thick membrane was patterned with metal stripes, featuring a period of 4 μm and a *duty cycle* of 80 %. As detailed in Extended Data Figure 1, this parameter choice follows from the need to tune the photonic resonance at the same energy of the intersubband transition.

We firstly characterized the sample by means of single-beam spectroscopy, hence accessing the usual polaritonic spectra. The single-beam absorbance is plotted in Fig. 2 (c) and shows that the first polaritonic peak (i.e. the only one accessible within the laser tuning range) is in good agreement with the theoretical curves obtained for the parameter set $\omega_0$ = 124.5 meV, $\gamma_r$ = 3 meV, $\gamma_{nr}$ = 0, $\gamma_m$ = 5 meV, $\Omega$ = 8 meV. Except for $\Omega$ and $\gamma_m$, these values were also confirmed by independent experiments on a photonic crystal bare resonator (transmission spectrum of a membrane with no active quantum wells) and on the plain, unpatterned quantum well heterostructure, as detailed in Extended Data Figure 1. The proximity of the absorption to the 50% value is very promising for the observation of CPA. As a matter of fact, the experimental points plotted in Figure 2 (c) are the average between the two single-beam absorbances, that slightly differ from each other in agreement with the differences in reflectance displayed in Figure 2 (d). This has to be attributed to the existence of a second resonant photonic mode, located at about 150 meV, which is not included in the CMT model introduced in this paper. A more accurate description of the spectra can be achieved by rigorous coupled-wave analysis (RCWA, dashed lines in the plots), where the structure is described *ab initio* starting from the geometrical parameters and the dielectric response of the semiconductor sample. The drawback of the RCWA model is that it does not provide any insight about the critical coupling in terms of damping rates matching. It has to be considered then as a complementary tool for a full interpretation of the experimental data. Incidentally, we note that the splitting between the polariton peaks in transmittance, reflectance and absorbance is not the same, as already pointed out in [22].

The results of the double-beam experiment are reported in Fig. 3, panels (a) – (c), where we plot the output intensities at ports 1 and 2 upon a sweep of the input beam phase difference $\varphi = \arg(s_2^+/s_1^+)$, for three different excitation wavelengths. In all three cases the total output intensity $|s_1^-|^2 + |s_2^-|^2$ reaches 2, i.e. CPT (units are chosen such that input intensities $|s_1^+|^2 = |s_2^+|^2 = 1$); at the wavelength corresponding to (b) the $\varphi$-phase sweep also enables to reach $|s_1^-|^2 + |s_2^-|^2 \approx 0$, hence CPA. The corresponding joint absorbance values are reported in panel (d) as a function of the wavelength for the three phases $\varphi$ corresponding to minimum, average, and maximum output. The CMT and RCWA theoretical traces are also shown for comparison. It is worth noticing that the average joint absorbance, measured via the double-beam experiment, closely follows the average single-beam absorbances (Fig. 2 (c)). The wavelength tuning of the laser covers basically the whole lower peak of the polariton doublet up to the resonance energy, unambiguously proving the polaritonic nature of the CPA and the existence of strong critical coupling. The occurrence of strong critical coupling in photonic crystal membrane intersubband polariton samples naturally follows from the very fact that in these samples the cavity and matter damping rates are close to each other. Indeed, it is a general feature following from the CMT that the modulation depth of the double-peaked absorption, and hence the strong critical coupling, is rather forgiving with respect to the relative damping rate mismatch. In our case, despite of $(\gamma_m - \gamma_r)/(\gamma_m + \gamma_r) = 25\%$, we observed $A_{\text{joint, max}} - A_{\text{joint, min}} \approx 90\%$. Further confirmation that the full coherent modulation of the absorption with phase difference is not an artifact due to other components in the setup is also given by the analysis of the output beam dephasing $\Delta\psi$. As detailed in the Supplementary Information, $\Delta\psi$ is connected to the phases of certain scattering-matrix elements, and hence represents a key feature of the sample under test. In panel (e) of Fig. 3 we report the measured $\Delta\psi$, and compare it with the CMT and RCWA calculations. Excellent agreement with the latter is observed, while the difference with the CMT can be again attributed to the presence of the second photonic mode at higher energy.

We conclude providing an experimental estimate of the S-matrix determinant. This estimate does not rely on the link between det $S$ and the absorbance (Eq. 3) – which is valid only in this CMT model – but rather on the general relation

$$\left|\det S\right| = \left|T - e^{i\Delta\psi}\sqrt{R_1 R_2}\right|.$$

Employing the single-beam $R$ and $T$, and the dephasing $\Delta\psi$ extracted from the double-beam experiment, we get the curve reported in Fig. 3 (e), which shows excellent agreement with the theory.

In summary, we have observed for the first time coherent perfect absorption and coherent perfect transparency (and hence complete interferometric control of absorption) by dressed light-matter states. The phenomenon is interpreted via a semiclassical model in terms of damping rates matching; this allowed us to generalize the critical coupling condition - already known for lossy resonators - to the strong-coupling polariton framework, with potential applications to all the situations where two coupled oscillators are driven by two coherent sources. Further studies are necessary to explore the true quantum regime, in which pairs of individual photons are now used to drive the strongly coupled cavity-emitter system, which - ultimately - could feature a single quantum emitter. A generalization of the Hong-Ou-Mandel dip phenomenology has been predicted in the case of coherent perfect absorbers [31], but this physics is still completely unexplored for polaritonic states.

**Acknowledgments**

We would like to thank G. Scamarcio and M. Liscidini for several fruitful interactions and discussions, E. Zanotto for providing us the loudspeaker actuator, and V. Spagnolo for the precious support with the laser source.

This work was supported in part by the Italian Ministry for Economic Development through the Teragraph project and by the European Research Council through the Advanced Grant SoulMan. R.C. acknowledges partial support from the ERC GEM grant (Grant Agreement No. 306661).

**Supplementary Material**

**Explicit expression of some quantities relevant in the coupled mode-theory.** The polaritonic poles appearing in Eq. (2) are

$$\omega_\pm = \omega_0 + \left[i(\gamma_r + \gamma_{nr} + \gamma_m) \pm \sqrt{4\Omega^2 - (\gamma_r + \gamma_{nr} - \gamma_m)^2}\right]/2,$$

while the determinant of the scattering matrix – apart from a phase factor - is given by

$$\det S(\omega) = \frac{(\omega - \varpi_+)(\omega - \varpi_-)}{(\omega - \omega_+)(\omega - \omega_-)}.$$

The determinant zeros are

$$\varpi_\pm = \omega_0 + \left[i(-\gamma_r + \gamma_{nr} + \gamma_m) \pm \sqrt{4\Omega^2 - (\gamma_r - \gamma_{nr} + \gamma_m)^2}\right]/2,$$

note the sign skew with respect to the poles. Real zeros, i.e. CPA, can be attained in two independent ways. The first one is when the argument of the square root is positive and the imaginary contribution involving the γ's is zero. This is the *strong critical coupling*, i.e. the situation where two zeros occur at

$$\tilde{\omega}_\pm = \omega_0 \pm \sqrt{\Omega^2 - \gamma_m^2}$$

under the condition $\gamma_r = \gamma_{nr} + \gamma_m$.

The second is when the argument of the square root is negative and the imaginary result compensates for the γ's; this is the *weak critical coupling*, i.e. the situation where one zero occurs at $\omega_0$ under the condition $\gamma_m (\gamma_r - \gamma_{nr}) = \Omega^2$.

**Two-port coherent (perfect) absorption, C(P)A.** The response of a linear system to an external drive can be described by a scattering matrix connecting input with output amplitudes: $|s^-\rangle = S |s^+\rangle$. Enforcing reciprocity, the most general $S$ can be written as

$$S = e^{i\vartheta} \begin{pmatrix} \rho_1 e^{i\psi_1} & i\tau \\ i\tau & \rho_2 e^{i\psi_2} \end{pmatrix}$$

where $0 \leq \rho_1, \rho_2, \tau \leq 1$. If $\det S = 0$, there exists a vector $|s^+\rangle$ for which $|s^-\rangle = |0\rangle$: this is CPA. Explicitly, the non-trivial solution requires $\rho_1 \rho_2 = \tau^2$ and $\psi_1 + \psi_2 = (2m+1)\pi$, where $m$ is an integer.

In general, if the system is excited from both ports with arbitrary amplitudes $|s^+\rangle = (s_1^+, s_2^+)$, the ratio (absorbed energy)/(input energy) is defined as *joint absorbance* $A_{\text{joint}}$. As the dephasing between the input beams $\varphi = \arg(s_2^+/s_1^+)$ is swept, the joint absorbance oscillates sinusoidally with φ, reaching a minimum and a maximum in one period of the optical dephasing. If the input beams have the same intensity ($|s_1^+|^2 = |s_2^+|^2$), the joint absorbance reaches minimum and maximum values given by

$$A_{\text{joint, min/max}} = (A_1 + A_2)/2 \pm A_{\text{mod}}$$

where $A_{1,2} = 1 - \rho_{1,2}^2 - \tau^2$ are the single beam absorbances, and the *modulating absorbance* is

$$A_{mod} = \sqrt{(1-A_1)(1-A_2) - |\det S|^2}.$$

Note that the average joint absorbance is the arithmetic mean of the single-beam absorbances. If the system is symmetric ($\rho_1 = \rho_2$, $\psi_1 = \psi_2$), det $S = 0$ implies $A_{joint} = 1$ (CPA); for an asymmetric system with det $S = 0$ the CPA is reached in general for different amplitude input beams. However, it can be shown that the effect of asymmetry is quadratic in ($A_1 - A_2$), hence an asymmetric sample having det $S = 0$ will anyway exhibit a very large joint absorbance even when driven with equal amplitude input beams.

While the individual phases $\psi_{1,2}$ can be experimentally accessed only if the absolute phase delay $\varphi$ is known, the phase *sum* $\psi_1 + \psi_2$ can be directly deduced by measuring the phase *difference* between output intensities at the ports 1 and 2 in the coherent absorption set-up. Indeed, if the inputs satisfy $|s_1^+|^2 = |s_2^+|^2 = 1$, the following expressions hold:

$$\left|s_1^-\right|^2 = \rho_1^2 + \tau + 2\rho_1 \tau \sin(\varphi - \psi_1 + \pi)$$

$$\left|s_2^-\right|^2 = \rho_2^2 + \tau + 2\rho_2 \tau \sin(\varphi + \psi_2).$$

Hence, by defining $\Delta\psi$ as the output beam phase difference, one gets $\Delta\psi = \psi_1 + \psi_2 - \pi$. Since all the prefactors in the above equations are positive, CPA can be attained only if the output beams are in phase, and the condition $\Delta\psi = 2m\pi$ coincides with the above stated requirement $\psi_1 + \psi_2 = (2m+1)\pi$.

**Figure 1. Coupled oscillator model, strong/weak critical coupling, and polaritonic CPA.** Panel (a): Two-port symmetric photonic resonator coupled to two scattering channels (ports) via the coupling constant $d$ (linked to the radiative damping rate $\gamma_r$), and to a matter resonator depicted as a spring-mass oscillator. Its linear response is given by the scattering matrix $S(\omega)$; when its determinant is zero all the incoming energy is absorbed (coherent perfect absorption, CPA). Panel (b), main graph: phase diagram of the coupled system. For small enough damping rates (red regions) the absorbance (subplots) exhibits two polaritonic peaks, while for large damping rates (orange regions) single-peaked spectra are observed. The dashed line marks the separation between the two regions. In both of these regions the joint absorbance can reach unity (CPA) provided that a *critical coupling* condition is fulfilled (see text). Panel (c): phase diagram with no cavity-matter coupling ($\Omega = 0$): the ordinary critical coupling condition $\gamma_r = \gamma_{nr}$ is recovered.

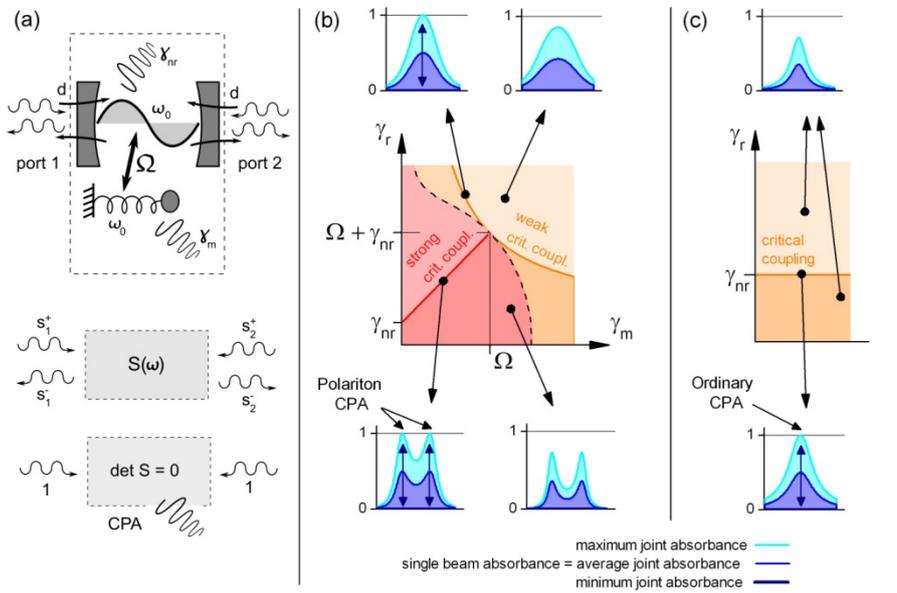

**Figure 2. Experimental set-up, sample details and single-beam spectra.** Panel (a): Interferometric arrangement for double-beam probing of the photonic crystal sample. The main features of the sample are outlined in panel (b): a thin semiconductor membrane is patterned with gold stripes implementing a metallic-dielectric photonic crystal resonator. In the membrane a multi-quantum well heterostructure acts as a collection of harmonic oscillators with resonance frequency $\omega_{12}$. Panels (c) and (d), large dots: experimental single-beam absorbance, reflectance and transmittance obtained by blocking one of the two interferometer's input beams. Solid lines represent the coupled-mode theory traces (CMT), while dashed lines are obtained via rigorous coupled-wave analysis (RCWA). The mismatch in reflectance and transmittance between CMT and experiment is possibly due to a second photonic resonance located at about 150 meV. This is also responsible for the difference between reflectances at ports 1 and 2. The transmittance was also measured with a broadband light source [small gray dots in panel (d)] to fully reveal the double-peak polaritonic structure; the perfect matching with the theoretical RCWA spectrum is here apparent.

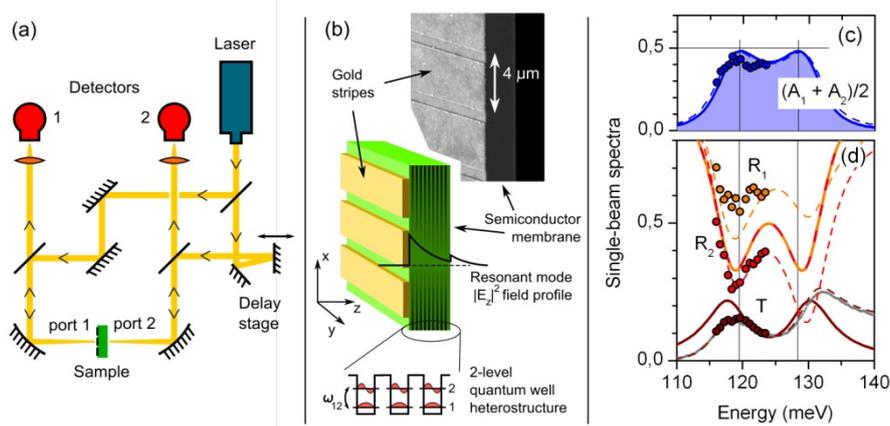

**Figure 3. Modulation of output intensity upon double-beam excitation, and polariton CPA.**
Panels (a-c): when the sample is excited from both ports 1 and 2 with unit-amplitude beams phase-shifted by φ, the output intensity recorded at ports 1 and 2 oscillates with a phase shift Δψ; the experimental traces (dots) are fitted by a sinusoid. Maximum, average, and minimum total output intensities are plotted in terms of joint absorbance [panel (d)], showing very good agreement with the two-peak polaritonic CPA theoretical traces (solid lines, coupled-mode theory; dashed lines, rigorous coupled wave analysis). Further confirmation of the phenomenon is gained by analyzing the S-matrix determinant and the output beam dephasing [panel (e)]. Panels (a)-(c) correspond to the energies labeled A-C in panels (d) and (e).

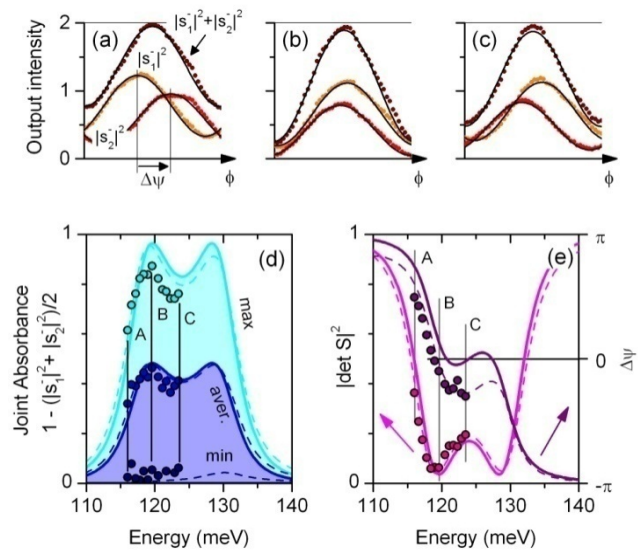

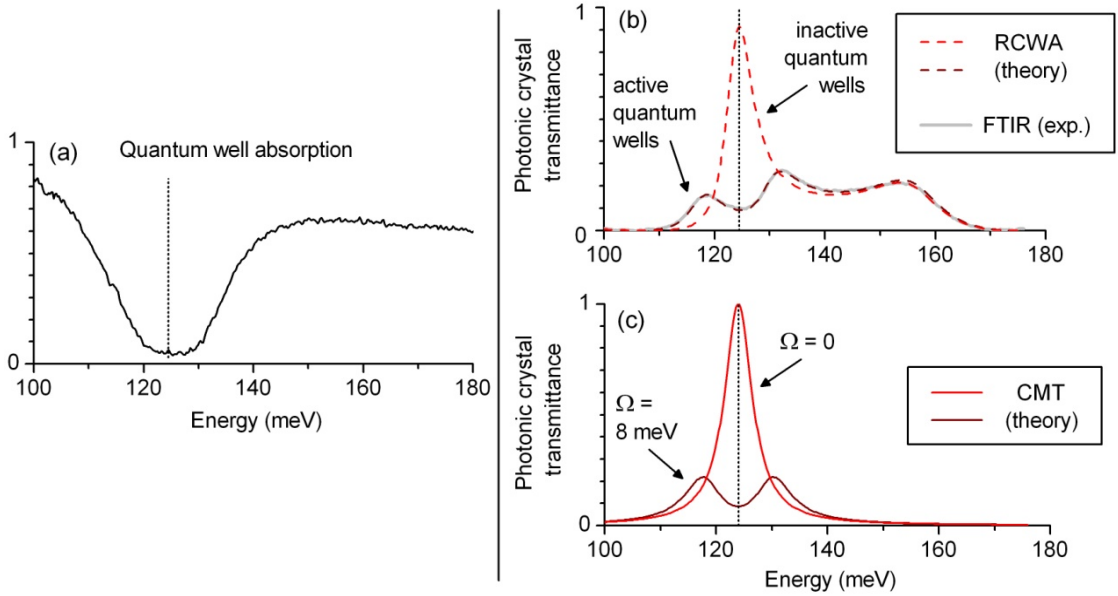

**Extended Data Figure 1. Overlook on the procedure for extracting the parameters employed in the models.** Panel (a): determination of the intersubband transition frequency $\omega_0 \approx 124$ meV, through a measurement of the quantum well (QW) absorption in the multipass configuration. Panel (b): Tuning the resonance of the photonic crystal, i.e. the optical cavity resonance, in order to match $\omega_0$, via a rigorous coupled wave analysis simulation (RCWA). In this simulation, the QWs are considered to be inactive, by setting to zero the charge density in the wells [30]. Meanwhile, from the width of the transmittance peak, the photonic cavity lifetime is retrieved ($\gamma_c = \gamma_r + \gamma_{nr} = 3$ meV). When introducing in the RCWA the nominal charge density n = $5 \cdot 10^{11}$ cm$^{-2}$, the double-peaked polaritonic spectrum pertaining to the sample with active QWs is obtained. A perfect agreement with the experimental spectrum measured by means of Fourier-transform spectrometry (FTIR) is obtained by fitting the FTIR transmittance with respect to the actual charge density, to the intersubband transition frequency, and to the intersubband transition decay rate. This procedure gave the parameters n = $5.2 \cdot 10^{11}$ cm$^{-2}$, $\omega_0 = 124.5$ meV, and $\gamma_m = 5$ meV. It should be noticed that the latter is not the half linewidth appearing in the QW multipass absorption measurement. In the multipass experiment, an artifact due to the saturation of absorption occurs, and neither the linewidth nor the lineshape can be trusted. This is because several, highly absorbing QWs are crossed by the probe light beam. Absorption measurements at the Brewster angle, performed on similar QW samples, showed unsaturated absorption lineshapes revealing the Lorentzian character of the intersubband transition broadening. In panel (b) we report the spectra obtained with the coupled mode theory (CMT). Here, the bare photonic cavity response (i.e. that for $\Omega = 0$, corresponding to the inactive QWs), is reproduced by inserting in Equation 2 of the main text the parameters $\omega_0 = 124.5$ meV, $\gamma_r = 3$ meV, $\gamma_{nr} = 0$. The last position is needed for reproducing a fully contrasted transmittance feature, and reveals that there are no non-resonant losses in the system. The spectrum, slightly differs from the corresponding one obtained by RCWA since the latter presents a feature at $\approx 155$ meV. This feature is attributed to a second photonic mode: while it appears in the experiment and in the full electromagnetic simulation (RCWA), it is not included in the coupled mode theory. Finally, by setting $\Omega = 8$ meV in the CMT, the double-peaked polaritonic spectrum is retrieved. The choice $\Omega = 8$ meV gives a good agreement of the absorption lineshapes reported in Figs. 2 and 3 of the main text, and is consistent with the usual expression linking the charge density with the Rabi splitting (see, e.g. Ref. [27]).